\documentclass[aps,pre,amsmath,amssymb,twocolumn,showpacs,byrevtex,superscriptaddress]{revtex4}
\usepackage{graphicx}
\begin{document}
\title{Energy landscape of a simple model for strong liquids}


\author{A. J. Moreno}
\affiliation{
         {Dipartimento di Fisica and INFM-CRS-SMC,
         Universit\`a di Roma {\em La Sapienza}, P.le. A. Moro 2, 00185 Roma, Italy}
        }
\author{ S. V. Buldyrev}
   \affiliation{
         {Yeshiva University, Department of Physics, 
	 500 W 185th Street New York, NY 10033, USA}
        }
\author{E. La Nave}
  \affiliation{
         {\hspace{-1 mm}Dipartimento di Fisica and INFM-CRS-SOFT,
         Universit\`a di Roma {\em La Sapienza}, P.le. A. Moro 2, 00185 Roma, Italy}
        }
\author{I. Saika-Voivod}
  \affiliation{
         {\hspace{-1 mm}Dipartimento di Fisica and INFM-CRS-SOFT,
         Universit\`a di Roma {\em La Sapienza}, P.le. A. Moro 2, 00185 Roma, Italy}
        }
\author{F. Sciortino}
  \affiliation{
         {\hspace{-1 mm}Dipartimento di Fisica and INFM-CRS-SOFT,
         Universit\`a di Roma {\em La Sapienza}, P.le. A. Moro 2, 00185 Roma, Italy}
        }
\author{P. Tartaglia}
\affiliation{
         {Dipartimento di Fisica and INFM-CRS-SMC,
         Universit\`a di Roma {\em La Sapienza}, P.le. A. Moro 2, 00185 Roma, Italy}
        }
\author{E. Zaccarelli}
  \affiliation{
         {\hspace{-1 mm}Dipartimento di Fisica and INFM-CRS-SOFT,
         Universit\`a di Roma {\em La Sapienza}, P.le. A. Moro 2, 00185 Roma, Italy}
        }

\begin{abstract}
We calculate the statistical properties of the energy landscape of a minimal model for 
strong network-forming liquids. Dynamic and thermodynamic
properties of this model can be computed with arbitrary precision even at low temperatures.
A degenerate disordered ground state and logarithmic statistics for the local minima 
energy distribution are the landscape signatures of strong liquid behavior. 
Differences from fragile liquid properties
are attributed to the presence of a discrete energy scale, provided
by the particle bonds, and to the intrinsic degeneracy of topologically
disordered networks.
\end{abstract}

\maketitle

In recent years, the study of the statistical properties
of the potential energy surface or landscape
(PES) \cite{wales,debenedetti01,angell95}  sampled by liquids
in supercooled states has received considerable attention, in the attempt
to quantify the thermodynamic properties of supercooled liquids and 
glasses \cite{stillinger_pes,sastry01,st,press}.
The PES is analyzed to estimate the number
and energy distribution $\Omega(E)$ of the local
minima and the  volume in configuration space sampled in vibrational
motions around each local minimum. 

According to Angell's classification \cite{fragdef}, a liquid is named
"fragile" if the temperature dependence of its transport coefficients shows large deviations 
from Arrhenius behavior. If no deviations are observed, the liquid is named "strong".
For the case of fragile liquids ---
in the region of energies sampled during low temperature $T$
equilibrium simulations with state of the art numerical resources ---
$\Omega(E)$ is well described by
a Gaussian distribution \cite{sastry01,heuer00,press}. The validity of
Gaussian statistics in regions of the PES with energy lower than those
accessible in simulations, would suggest the existence of a finite Kauzmann
temperature $T_K$, where the configurational entropy $S_{conf}$ vanishes,
and via the Adam-Gibbs relation \cite{adam65,wolynes}, a
divergence of the characteristic relaxation time at $T_K$. The
quantification of the statistical properties of the PES for models of
strong liquids is still under debate \cite{voivod01,newheuer}. It has
been shown that, on lowering $T$, deviations from Gaussian
statistics take place. The configurational entropy  does not seem to
extrapolate to zero at a finite $T$, but the long equilibration times
and the unknown value of the  ground state energy prevent an
unambiguous determination of the ground state degeneracy. A recent
work of Heuer and coworkers \cite{newheuer} suggests that the breakdown
of Gaussian statistics originates from the emergence of a
natural cut-off in $\Omega(E)$, related to the formation of a fully connected network of
bonds.

In this Letter, we report a study  aiming to clarify the statistical
properties of the PES for a strong liquid, to provide a useful
framework for interpreting results of realistic models, and to shed
light on the differences between fragile and strong liquids. We
quantify the landscape properties for a simple model, similar in spirit to one 
previously introduced by Speedy and Debenedetti \cite{speedymaxval}. We
show that, with appropriate numerical techniques, both the dynamics and 
thermodynamic properties of this model can be computed with high accuracy
and that no extrapolations are required to determine the low $T$ behavior.
The strong liquid behavior is associated with the presence
of two ingredients, which are characteristic of all network-forming
liquids: a degenerate ground state and a discrete distribution of energies
above the ground state, generated by the bond energy scale.

We investigate a maximum valency model: a square well
model of width $\Delta$ with a constraint on the maximum number of interacting
particles. The interaction between two particles $i, j$ that each have
less than $N_{max}$ bonds to other particles, or between two particles
already bonded to each other, is given by a square-well potential,
\begin{equation}
V_{ij}(r)=
\begin{cases}
	~~\infty ~~~~~~~\hspace{0.8 mm} r<\sigma    \\ 
        -u_0  ~~~~~~~ \sigma<r<\Delta  \\
        ~~~0      ~~~~~~~~ r>\sigma+\Delta.	
\end{cases}
\label{eq:model1}
\end{equation}
When $i$ and/or $j$ are already bonded to $N_{max}$ neighbors,
then $V_{ij}(r)$ is simply a hard sphere (HS) interaction,
\begin{equation}
V_{ij}(r)=
\begin{cases}
	~~\infty ~~~~~~~~r<\sigma    \\ 
        ~~~0      ~~~~~~~~\hspace{0.8 mm} r>\sigma.	
\end{cases}
\label{eq:model2}
\end{equation}
The maximum number of bonds per particle is controlled by tuning
$N_{max}$. This model is particularly suited for theoretical and
numerical studies. First of all, along an isochore, the system changes from a HS
system (when $k_{B}T \gg u_0$, $k_{B}$ is the Boltzmann constant) to a simple
model for network-forming liquids (when $k_{B}T< u_0$) with coordination
$N_{max}$ and no angular constraints \cite{noangcons}.   
Second, the ground state energy
for a system of $N$ particles is known, being the energy of a fully
connected network ($-u_0 N N_{max}/2$). Third, as $V(r)$ is simply
a square well model, the local minima of the PES coincide with the
bonding patterns of the system. Consequently, moving between minima
takes place via breaking and reforming of bonds. Fourth, the volume
in configuration space of each of these bonding patterns can be
calculated with no approximation, as discussed in the following. In
the PES language, it means that the basin free energy \cite{stillinger_pes} can be
calculated exactly. Finally, since the energy levels of the system are discrete by
construction, the number of bonding patterns with $M<<N N_{max}/2$
broken bonds can be calculated by combinatorial factors.
Exploiting these properties allows us to quantify precisely the statistical
properties of the PES of this model.

\begin{figure}
\vspace{-1. mm}
\includegraphics[width=.27\textwidth]{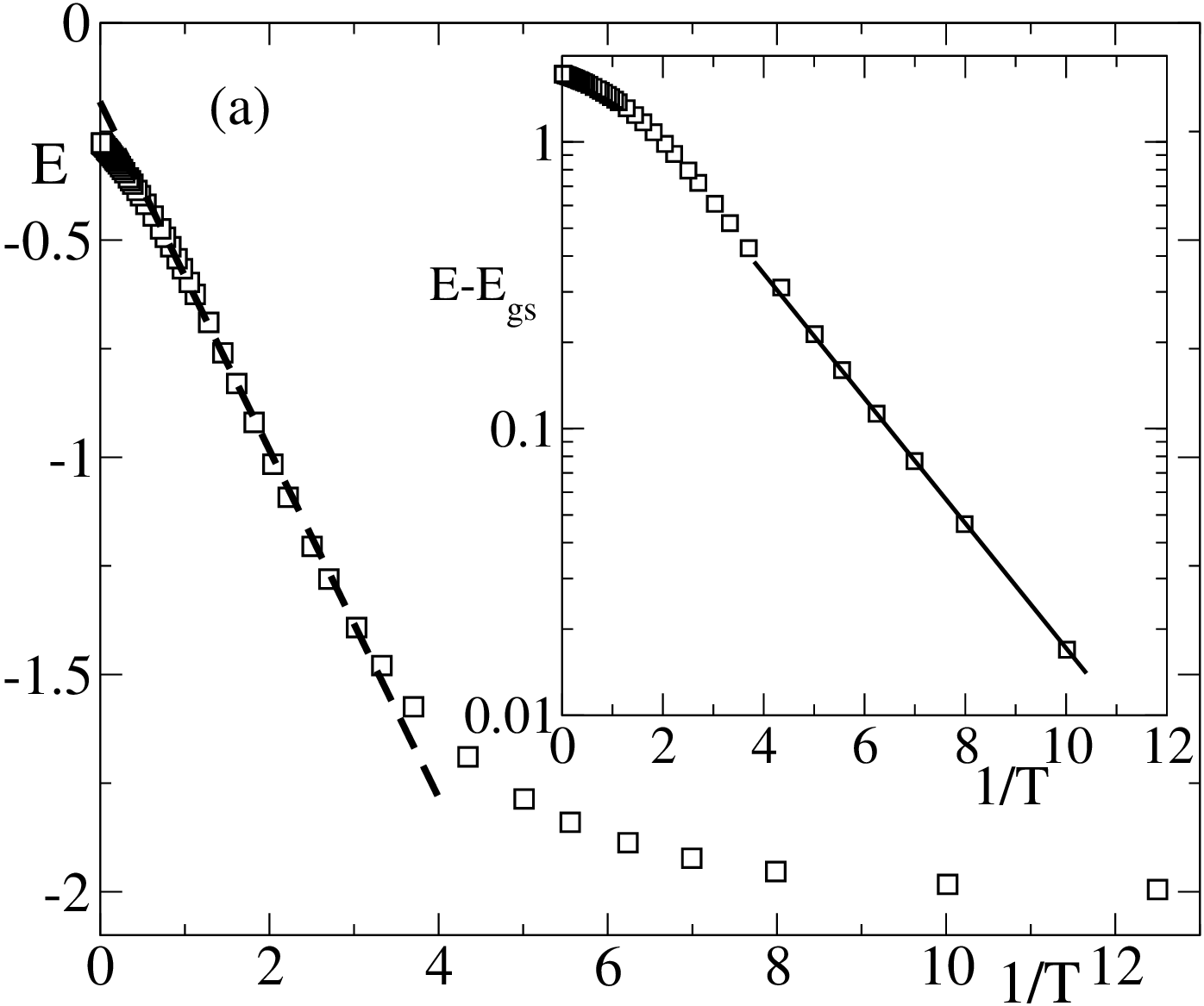}
\includegraphics[width=.27\textwidth]{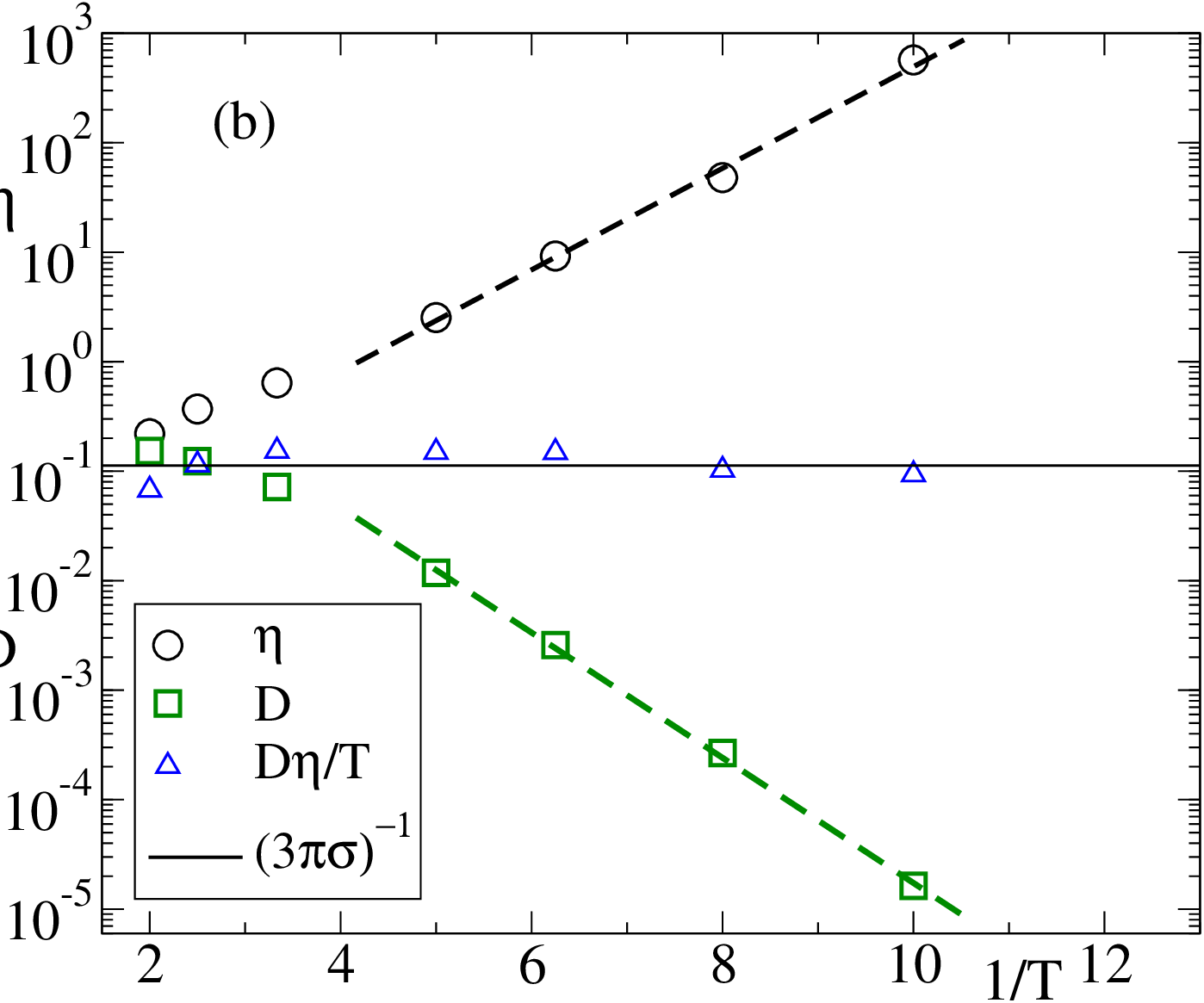}
\caption{Temperature dependence of 
the potential energy per particle $E$ (top) and of $\eta$
(in units of $(mu_{0})^{1/2}/\sigma^{2}$), $D$ (in units of $\sigma(u_{0}/m)^{1/2}$),
and the product $D\eta/T$  (bottom). $D$ is calculated from the long time limit of
$\delta r^2(t)/6t$, with $\delta r^{2}(t)$ the mean squared displacement, and $\eta$ as explained
in \cite{alder}. In the top panel,
the dashed line shows the $1/T$ behavior expected for a Gaussian landscape \cite{sastry01,REM}.
The inset shows the same data in a semi-log representation to highlight the
validity of the low-T behavior $\ln[E-E_{gs}]= \ln[2.56]-\frac{1}{2T}$
indicated by the continuous line (see text). In the bottom panel the dashed lines
are the best-fit Arrhenius behavior for $\eta$ and $D$. The continuous line corresponds to the expected
value from the Stokes-Einstein relation (see text).
\vspace{-2 mm}} 
\label{fig:e}
\end{figure}

We perform Monte Carlo (MC) and event driven molecular dynamics simulations
by using $\Delta/(\sigma+\Delta)=0.03$, $u_0=1$, $\sigma=1$ and $N_{max}=4$.
Entropy, $S$, is measured in units of $k_B$.  
Setting $k_B=1$, energy $E$ and temperature $T$ are measured in units of $u_0$. 
We study a system of $N=10^{4}$ particles of equal mass $m=1$,
implementing periodic boundary conditions,
at a fixed packing fraction $\phi=0.30$ \cite{packfrac} 
from temperature $T=100$ (where the HS limit is recovered)
down to $T=0.04$, where an almost fully connected network
of bonds is obtained \cite{mapping}. No evidence of phase separation
is observed \cite{zaccar}.
Fig. \ref{fig:e}a shows the
$T$ dependence of the potential energy per particle $E$ as a function of $1/T$.
For a Gaussian PES \cite{sastry01,REM}, $E$ should be linear in $1/T$. Deviations from the
$1/T$ law are clearly seen at low $T$, when the
energy approaches the ground state energy $E_{gs}=-2$, with a striking similarity with the
$T$ dependence of the energy of the sampled minima evaluated in landscape analyses of
atomistic models for silica \cite{voivod01,newheuer}.
The approach to $E_{gs}$ is well described by $E-E_{gs} \sim e^{-\frac{1}{2T}}$, providing
an unambiguous way of calculating all thermodynamic properties down to $T=0$.
  
Next we provide evidence that the present model is a simple and satisfactory one
for strong glass-forming liquids, by showing several features that
are commonly observed in real systems.
Fig. \ref{fig:e}b shows the $T$ dependence of $\eta$ and $D$ \cite{visc}.  
Both quantities display Arrhenius behavior at low $T$, as
expected for strong liquids, with an activation
energy controlled by the bond energy. The product
$\frac{D \eta}{T} \approx  \frac{1}{3\pi \sigma}$, as expected 
from the Stokes-Einstein relation. 
We also evaluate the exponent $\beta$ of the stretched exponential function that describes
the long time decay of density autocorrelation functions, finding values
$\beta > 0.85$ for all the wavevectors. Such a large $\beta$ value is in agreement with the
experimentally observed direct correlation between $\beta$ and strong behavior \cite{bohmer}. 
Finally, since at the investigated low $T$ the potential energy has already approached the ground state
value (Fig. \ref{fig:e}a), no signifficant drop in the specific heat will take place
at the glass transition temperature. This feature is often found in strong liquids \cite{martinez,richet}.

Next we evaluate the statistical properties of the PES. In
the landscape approach, configuration space is partitioned into basins around
the local minima of the PES. For the
present model, different local minima can be unambiguously identified
as different bonding patterns and the energy of the minimum coincides with the potential energy of the 
configuration, i.e. with the number of bonds.  The partition function can be written as
a sum over all bonding patterns. At low $T$, it is convenient
to write the sum over the number of $M$ broken bonds, associating with
each energy level $E$ the number of distinct network configurations with
$M \equiv (E-E_{gs})N $ broken bonds (the degeneracy of the energy level) times the volume
in configuration space that can be sampled by the network without
breaking or forming any bond (i.e. preserving the
bonding pattern). The logarithm of the number of distinct networks with the same number of bonds
defines $S_{conf}(E)$, while
the logarithm of the sampled volume defines the basin vibrational
entropy $S_{vib}$. The sum $S_{conf}+S_{vib}$ defines the total entropy of the model.

\begin{figure}
\includegraphics[width=.26\textwidth]{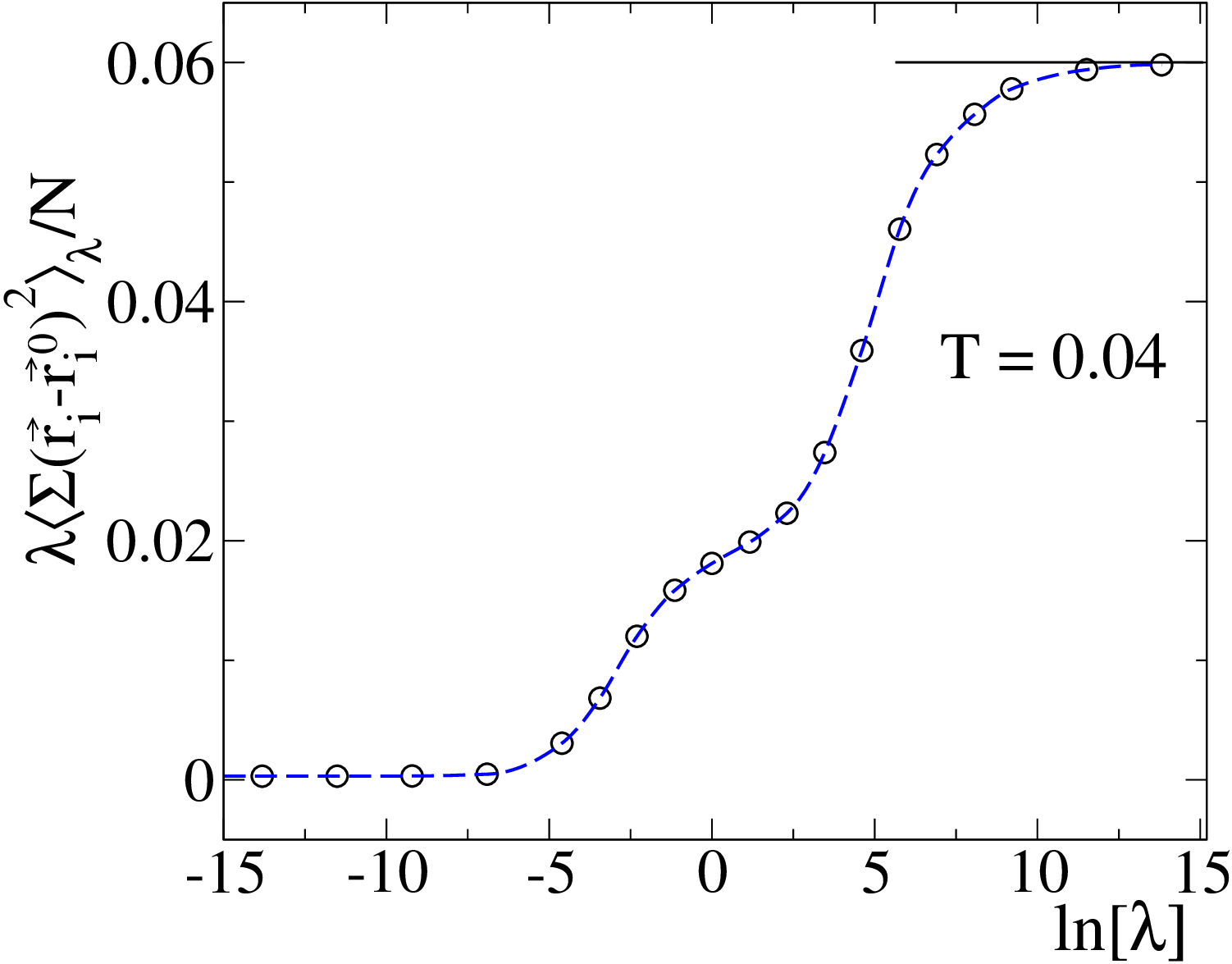}
\caption{Mean squared displacement per particle times $\lambda$
as a function of $ln[\lambda]$ for a typical basin
explored at $T=0.04$.  The continuous line  shows the harmonic behavior $ \frac{3}{2} k_{B}T$.
\vspace{-3 mm}}
\label{fig:ec}
\end{figure}

In the present model, $S_{vib}$ can be calculated without approximation by thermodynamic
integration from a reference Einstein crystal \cite{frenkel}. 
In this technique, a series of MC simulations
are performed adding to the original
Hamiltonian $H_0$ a harmonic perturbation 
$H_E(\vec r^N ;\lambda)=  \lambda \sum_{i=1}^{N} (\vec r_i - \vec r_i^0)^2 $,
where $(\vec r_1^0,...,\vec r_N^0)$ is a typical configuration of energy $E$  whose basin free energy
needs to be evaluated, and the parameter $\lambda$ is the elastic constant
of the harmonic perturbation. The system described by the 
Hamiltonian $H(\vec r^N;\lambda)= H_{0}(\vec r^N) +H_E(\vec r^N; \lambda)$ is then simulated
at fixed $T$ and $V$ with a MC technique, rejecting all moves altering
the bonding pattern. We study several values of $\lambda$ running from
$\lambda=0$ up to a value $\lambda_{max}$, whereupon the behavior of a
system composed of 3$N$ independent harmonic springs of elastic constant
$\lambda_{max}$ is recovered. In this limit, the free energy is known
analytically. The vibrational free energy $f_{vib}$
($\lambda=0$) of the basin to which $\vec r^{ 0}$ belongs is given by
\begin{eqnarray}
f_{vib}(T)= F_{H_E}(T;\lambda_{max}) \nonumber\\
 -\int_{-\infty}^{ln[\lambda_{max}]} \lambda 
 \langle \sum_{i=1}^{N} (\vec r_i - \vec r_i^{ 0})^2 \rangle_{\lambda} d ln[\lambda], 
\label{eq:finteg}
\end{eqnarray}
where $ \sum_i (\vec r_i - \vec r_i^{ 0})^2 $ is the sum of the
square displacements of all particles at a fixed value of
$\lambda$, brackets denote ensemble average, and $F_{H_E}(T;\lambda_{max})$ is the free energy of the 3$N$
Einstein oscillators.  Fig. \ref{fig:ec} shows the $\lambda$ dependence
of $ \lambda  \langle \sum_i (\vec r_i - \vec r_i^0)^2 \rangle_{\lambda} /N $ for one
specific state point. At large  $\lambda$ values, the mean square displacement 
per particle approaches the theoretically expected limit
$ \frac{3}{2}\frac{k_{B}T}{\lambda}$.  It must be noted that in the present model, since no
energy changes are involved in sampling the configuration space volume
at fixed bonding pattern, for each specific basin the basin free energy is purely entropic.  

\begin{figure}
\includegraphics[width=.255\textwidth]{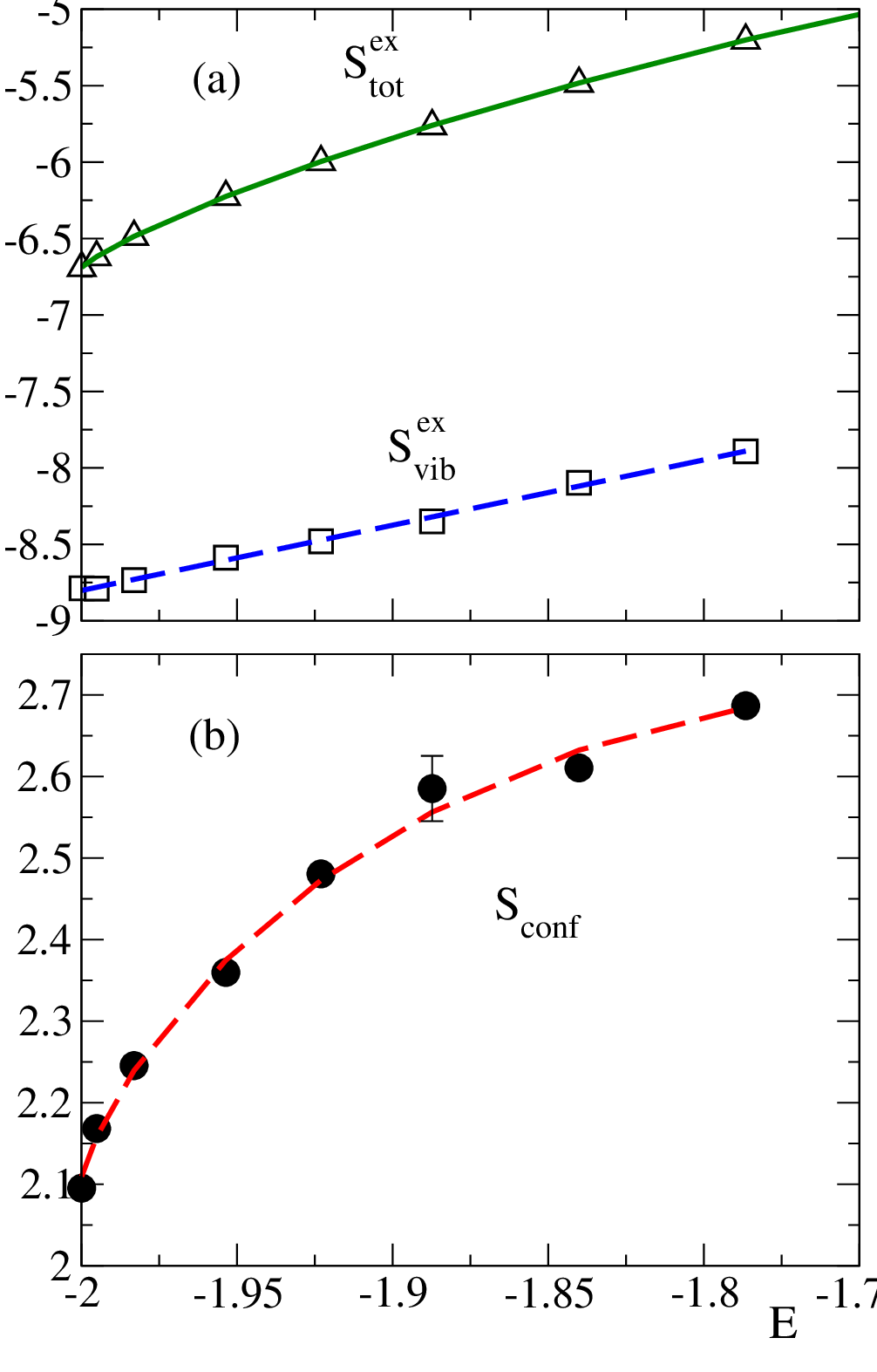}
\includegraphics[width=.255\textwidth]{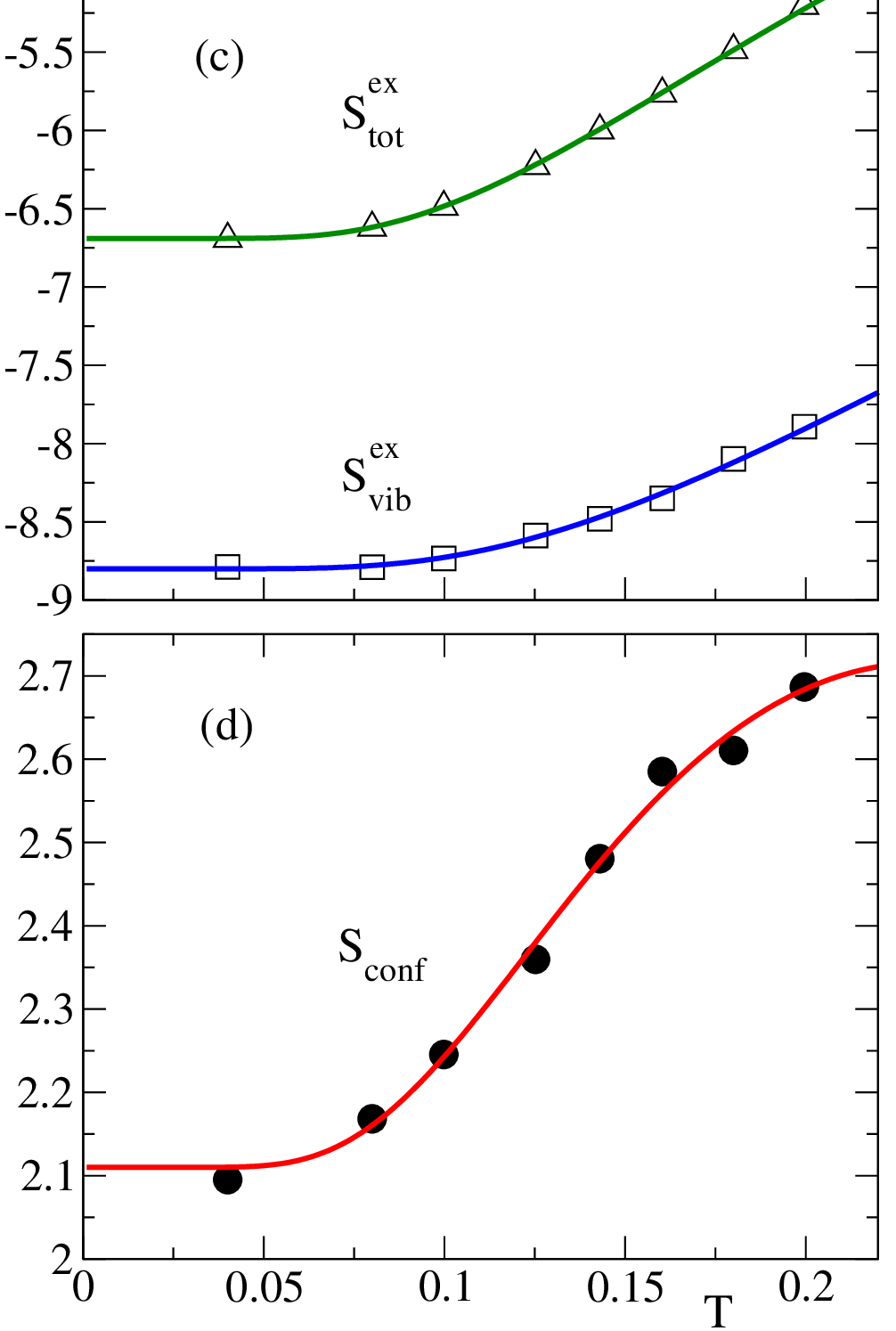}
\vspace{1 mm}
\caption{ Energy $E$ [panels (a) and (b)] and $T$ [panels (c) and (d)]  dependence of 
$S_{tot}^{ex}$, $S_{vib}^{ex}$ and $S_{conf}$.
A typical error bar is shown in panel (b).
Dashed lines for $S^{ex}_{conf}(E)$ and $S^{ex}_{vib}(E)$
are fits to Eqs.\ref{eq:fitsvibex}-\ref{eq:fitsconf} (see text).  
The continuous line for $S_{tot}^{ex}$ is the result of summing the obtained fit functions.
The continuous lines for $S_{tot}^{ex}(T)$, $S_{vib}^{ex}(T)$ and $S_{conf}(T)$
are calculated from the $T$ dependence of $E$ (see text).
\vspace{-4 mm}
}
\label{fig:s}
\end{figure}

Repeating the thermodynamic integration procedure for several different
equilibrium configurations at different $T$, $S_{vib}(E)$ can be precisely calculated.
Fig. \ref{fig:s}a shows that the excess quantity, $S_{vib}^{ex}(E)$, over the ideal gas contribution,  
depends linearly on the basin depth $E$, a feature 
shared with previously investigated models for supercooled liquids \cite{sastry01,press,wales}.
$S_{vib}^{ex}(E)$ is well represented by the best-fit function:
\begin{equation}
S^{ex}_{vib} = -8.8 + 4.27(E-E_{gs}).
\label{eq:fitsvibex}
\end{equation}
Complementing these data with the calculation of the excess total entropy over the ideal gas contribution,
$S_{tot}^{ex}$, by integration of the $T$ dependence
of the specific heat over $T$ from the known HS high $T$ reference state \cite{frenkel},
allows us to evaluate $S_{conf}\equiv S_{tot}^{ex}-S_{vib}^{ex}$.  
Fig.\ref{fig:s}b shows that, close to the ground state, the $E$ dependence of $S_{conf}$ is well described
by a logarithmic combinatorial function:
\begin{eqnarray}
S_{conf}(E) = S_{conf}(E_{gs}) \nonumber\\
 -2(E-E_{gs})\ln[2(E-E_{gs})]+(E-E_{gs}).
\label{eq:fitsconf}
\end{eqnarray}
The logarithmic term results
from the number of distinct ways $(E-E_{gs})N$ bonds can be broken among $N$ particles.
A one-parameter fit (see Fig. \ref{fig:s}b) suggests $S_{conf}(E_{gs}) \approx 2.1$ \cite{uppersconf}.  
Eqs. \ref{eq:fitsvibex}-\ref{eq:fitsconf} fully define the statistical properties of the PES.
All thermodynamic functions can be evaluated from them. Indeed, solving
$d(S_{conf}+S^{ex}_{vib})/dE = 1/T$, the $T$ dependence of $E$ is found to be
$E(T)-E_{gs}=2.56 e^{-\frac{1}{2T}}$, providing the continuous line in the inset of Fig. \ref{fig:e}a,
and parametrically, the continuous lines for $S^{ex}(T),S^{ex}_{vib}(T)$ and $S_{conf}(T)$,
compared with the numerical estimates in Figs. \ref{fig:s}c-\ref{fig:s}d.

According to the picture emerging from this study, strong liquid behavior
is connected to the existence of an energy scale (provided by
the bond energy) which is discrete and dominant as compared to the
energetic contributions coming from non-bonded next-nearest neighbor
interactions \cite{angellbond}. It is also intimately connected to the existence 
of a significantly degenerate ground state, favoring the
formation of highly bonded states which can still entropically
rearrange to form different bonding patterns with the same energy \cite{schwabl}. 
The presence of the bond energy scale \cite{impure} also determines a distribution of
energy levels above the disordered ground state which are
logarithmically spaced. These results help
rationalize previous landscape analysis of realistic models of
network-forming liquids \cite{voivod01} and the recent observation by Heuer
and coworkers that the breakdown
of Gaussian landscape statistics is associated with
the formation of a fully connected defect-free network \cite{newheuer}. 
While in realistic models the ground state energy is not known,
and the very long equilibration times needed
prevent an unambiguous determination of the
ground state degeneracy, both these quantities can be calculated for the
present model.

In this model, the discrete energy states provide a truly
degenerate ground state. In real systems, small differences
in energy exist, e.g. resulting from second neighbor interactions,
between configurations that in our model would correspond to
the same energy state. Hence, in real systems the ``ground state''
energy is smeared out, singling out a unique lowest energy state.
Consequently, for $k_BT$ values smaller than this spread in energy,
$S_{conf}$ approaches zero. Therefore, our {\it classical} model is strictly
applicable at temperatures where
quantum effects are negligible \cite{quantum}, and for values of $k_{B}T$
much larger than the spread in energy of the ground state.

The simplicity of the model and the possibility of calculating accurately
the relevant thermodynamic properties
at all $T$  makes this model a relevant
candidate for deepening our understanding of the differences between
strong and fragile liquids. Results reported here suggest
that strong and fragile liquids are characterized by significant
differences in their PES properties. In a simplified picture, 
a non-degenerate disordered ground state
and Gaussian statistics characterize fragile
liquids, while a degenerate disordered ground state and logarithmic statistics are associated
with strong liquids. The origin of this difference is traced
to the presence of a discrete energy scale, provided by particle bonds,
and to the intrinsic degeneracy of topologically disordered networks.

We acknowledge support from MIUR-COFIN, MIUR-FIRB, DIPC-Spain,
NSF and NSERC-Canada.  
%
%


%


\begin{thebibliography}{99}
\bibitem{noangcons}  In Ref.\cite{speedymaxval} 
an angular constraint is imposed by avoiding three particles bonding loops.
\bibitem{packfrac} 
Note that a diamond structure of touching hard spheres has $\phi \approx 0.34$, a value
compatible with the one we study here. The diamond structure is characteristic of
tetrahedral network forming strong liquids like water and silica.
\bibitem{mapping} 
In the effort to capture the basic ingredients of a strong liquid, angular correlations are
completely neglected. Still, the investigated low $T$-regime qualitatively corresponds to typical
experimental values. From SiO$_{2}$ and GeO$_{2}$ viscosity data above
the glass transition temperature \cite{richet,sipp}, we estimate Arrhenius activation
energies $E_A$(SiO$_{2}$) $\approx 62000$ K and
$E_A$(GeO$_{2}$) $\approx 31000$ K. Assuming $u_0 \approx E_A$ the investigated
$T$-range in our simulations roughly corresponds to $T > 2500$ K for
SiO$_{2}$ and $T > 1200$ K for GeO$_{2}$, 
values in the experimental accesible $T$-range \cite{richet,sipp}.
\bibitem{visc} Note that in simulations, it is today possible
to follow changes in dynamic properties covering up to six orders of magnitude,  
to be compared to changes in experimental quantities covering about 12 decades.
\bibitem{uppersconf} The value found here should be considered an upper bound value for
tetrahedral networks \cite{speedymaxval,sceats}, due to the absence of angular 
constraints in the bonding geometry.
\bibitem{impure} Impure network-forming liquids as mixed alkali silicates
\cite{richet,bohmer} show deviations from strong behavior. 
Such deviations could originate from the interplay between
different bond energies, strongly smearing out the discrete energy scale.
\end{thebibliography}
\end{document}